\documentclass[trackchanges,twocolumn]{aastex631}

\makeatletter
\let\frontmatter@title@above=\relax
\makeatother

\usepackage{url}
\usepackage{amsmath}

\usepackage{ amssymb }
\usepackage{appendix}
\usepackage{float}

\shorttitle{The Effect of a Dark Matter Core on the Structure of a Rotating Neutron Star}

\accepted{by ApJ, April 28, 2024}

\graphicspath{Figures/}
    \usepackage{graphicx}
    \usepackage{capt-of}
    \usepackage{lipsum}

\begin{document}
\shortauthors{Konstantinou 2024}
\correspondingauthor{Andreas Konstantinou}
\email{andreas.konstandinu@gmail.com}
\author[0000-0002-1072-7313]{Andreas Konstantinou}
\affiliation{Lympia, 2566, Nicosia, Cyprus}

\title{The Effect of a Dark Matter Core on the Structure of a Rotating Neutron Star}

\begin{abstract}
Neutron stars represent unique laboratories, offering insights into the physics of supranuclear-density matter and serving as potential hosts for dark matter. This study explores the impact of dark matter cores on rapidly rotating neutron stars through the two-fluid approximation, assuming minimal interaction between baryonic matter and dark matter. The investigation employs phenomenological models for fermionic and bosonic dark matter, revealing that universal relations governing mass and radius changes due to rotation remain largely unaffected in the presence of a dark matter core. Specifically, for a 5\% dark matter mass fraction, the percent deviations in total mass ($M_{tot}$), the baryonic equatorial radius ($R_{Be}$), and polar-to-equatorial baryonic radius ratio ($R_{ratioB}$) are within 3.9\%, 1.8\%, and 1.4\%, respectively. These findings suggest that the universal relations governing neutron star shape can be utilized to infer constraints on the properties of dark matter cores even in cases where the dark matter significantly softens the neutron star's equation of state. 
\end{abstract}

\keywords{neutron stars, relativistic stars, pulsars, stellar rotation, compact objects, millisecond pulsars, dark matter}

\section{Introduction} \label{sec:intro}

Neutron stars, with their dense cores and strong gravitational fields, are unique astrophysical laboratories for exploring the physics of the unknown equation of state (EOS) of cold, supranuclear-density matter. If the masses and radii of many neutron stars are measured, it would be possible to determine the EOS \citep{1992Lindblom}. However, the high-density interior may also serve as a site for non-standard model physics, leading to an accumulation of dark matter (DM) \citep{PhysRevD.40.3221}, which introduces significant changes to the possible masses and radii of the Dark Matter Admixed Neutron Stars (DMANS) \citep{LI201270,X.Y.Li_2012}.

Despite decades of research, the nature of dark matter, constituting a significant portion of the universe's mass-energy content, remains one of the most pressing questions in modern physics and cosmology (for a comprehensive review, see \cite{Bertone_2005}). Proposed candidates, from weakly interacting massive particles (WIMPs) \citep{Jungman_1996} to axions \citep{Sikivie_2010}, continue to elude direct detection. An intriguing avenue for indirect detection is introduced by considering neutron stars as potential hosts for compact fermionic or bosonic dark matter structures. Considerable work has been devoted to exploring the structure of non-rotating DMANS.  For instance,  DMANS have been constructed using mirror baryonic DM \citep{SANDIN2009278}, non-annihilating DM particles \citep{CIARCELLUTI201119,PhysRevD.84.107301},  self-interacting fermionic asymmetric DM \citep{GOLDMAN2013200,PhysRevD.92.063526,Mukhopadhyay_2017,Miao_2022},  fermionic dark matter particles (interacting with each other by exchanging mesons) \citep{PhysRevC.89.025803}, fermionic dark matter particles (interacting with each other by exchanging Standard Model Higgs bosons) \citep{PhysRevD.96.083004}, bosonic DMANS \citep{PhysRevD.105.023001}, light dark matter \citep{Motta_2018}, 
and a DM EOS that is obtained from the rotational curves of galaxies \citep{Rezaei_2017}.

A challenge arises from the fact that the EOS for dense-cold nuclear matter remains unknown, introducing additional uncertainty into the computation of the properties of DMANS. For instance, the DMANS' mass-radius relation depends on both the chosen EOS for the neutron star and the dark matter. 

Each candidate baryonic EOS uniquely determines a mass-radius curve for a neutron star, so measurements of the masses and radii of many neutron stars can constrain the neutron star's  EOS. One method for measuring the mass and radius is through observations of pulsed x-ray emission from millisecond-period rotation-powered pulsars \citep{2016Watts}, as is done by the NICER x-ray telescope \citep{2012SPIE.8443E..13Gendreau}.

The standard techniques employed in the analysis of NICER data assume that there is no dark matter present in any of the pulsars observed. But recently, \citet{2023PhRvD.107j3051R} have proposed an analysis method that would make use of NICER observations to provide simultaneous constraints on both the baryonic and dark matter EOSs, assuming the pulsars are DMANS. Their work is restricted to the cases when all of the dark matter is confined within the baryonic surface of the DMANS, a configuration called a dark matter core. 

Depending on the properties of the dark matter particles, theoretical models predict that dark matter may extend beyond the baryonic surface, creating what is known as a dark matter halo. This extended configuration has the potential to influence the X-ray pulse profile of the DMANS \citep{Miao_2022}. 

The pulse-profile modelling method used to analyze NICER data is sensitive to the oblate shape of the rotating neutron stars. The shape of the rotating star for pure neutron stars is described by a simple universal formula that only depends on the mass, equatorial radius, and the angular velocity, allowing computationally inexpensive parameter estimations \citep{2019ApJ...887L..21R}. In order to analyse NICER data for DMANS, it is important to understand the properties of rotating DMANS and to determine if the universal relation for the neutron star's shape still holds when dark matter is present inside the star.

In this paper the structure of rapidly rotating neutron stars with dark matter cores are computed using the two-fluid approximation. In the two-fluid approximation, it is assumed that the baryonic matter and the dark matter only interact through gravity. The properties of the rotating DMANS will be used to investigate departures due to dark matter cores from the universal relations for the increase in mass and radius due to rotation \citep{Konstantinou_2022} and the oblate shape due to rotation \citep{2014AlGendy}.

Prior research has delved into the structure of a two-fluid rotating DM-admixed white dwarf star in the framework of Newtonian gravity \citep{2022ApJ...941..115C,2023ApJ...945..133C}. 
A similar situation is that of superfluid neutron stars, which are computed using a weakly interacting two-fluid formalism in general relativity for slowly rotating \citep{2002A&A...381..178Prix2002} and rapidly rotating  \citep{2005PhRvD..71d3005Prix} neutron stars.
 An alternative approach for rapidly rotating DMANS assumes a weak interaction between dark and standard sectors \citep{2021JCAP...09..027G}.

There are a few computations of rapidly rotating DMANS in the two fluid approximation.  \citet{Mukhopadhyay_2017}, solved the Einstein's field equations (EFE) using the \texttt{LORENE} code \citep{gourgoulhon2011introduction}. In this paper the EFE equations are solved using the same methods introduced by \citet{1989Komatsu,1992Cook} using a modification of the \texttt{rns} code \citep{1995Stergioulas} to the two-fluid formalism. An independent modification of  \texttt{rns} has been implemented by by \citet{2024arXiv240313052C}, where the situation of a fermion-boson star has been studied.

This paper is organized as follows. In section \ref{sec:EOS}, the equations of state for fermionic and bosonic dark matter and nuclear matter are introduced in accordance with the methods outlined in other papers. In section \ref{sec:2fluid} the structural equations of a two-fluid star are reviewed. Subsection \ref{sec:2TOV} provides an overview on how the Tolman-Oppenheimer-Volkoff (TOV) equations \citep{1939Oppenheimer,1939Tolman} are modified when considering two non-interacting fluids. In subsection \ref{sec:2rns}, the structure equations for a two-fluid rapidly rotating star and the modified \texttt{rns} code are introduced.
In Section \ref{sec:Univ}, the universal relations governing the oblate shape and the change of mass and radius due to rotation are revisited. 
In Section \ref{sec:non-universal}, the main findings are presented. Specifically, it is assumed that the frequency of the DM fluid is zero, and the baryonic fluid can rotate rapidly. It is observed that the universal relations discussed in Section \ref{sec:Univ} remain largely unaffected by the presence of a dark matter core.
 The conclusion and the initiation of a discussion are conducted in Section \ref{sec:discussion}.

The unit system chosen is one in which $\hbar$, c, and G are set to 1.

\section{Dark Matter and Neutron Star Equation of State}\label{sec:EOS}
\subsection{Fermionic Dark Matter}\label{sec:ferm}
The nature of DM remains unknown. Therefore, there is ample room for experimenting with interesting concepts. One such concept is the assumption of cold, self-interacting, fermionic DM. These particles are non-annihilating, as in asymmetric DM \citep{ZUREK201491}. The energy density of a system with particles having mass, number density, and kinetic energy density equal to $m_\chi$, $n_\chi$, and $\epsilon_{kin}$, respectively, is given by \citep{PhysRevD.74.063003,Nelson_2019}
\begin{equation}
    \epsilon_\chi=\epsilon_{kin}+m_\chi n_\chi + \frac{ n^2_\chi}{m^2_I},
\end{equation}
where
\begin{equation}
    \epsilon_{kin}=\frac{1}{\pi^2}\int_0^{k_F} dkk^2(\sqrt{k^2+m^2_\chi}-m_\chi),
\end{equation}
with $k_F = (3\pi^2 n_\chi)^{1/3}$. $m_I$ describes the repulsive interaction energy scale among the DM particles \citep{PhysRevD.92.063526,PhysRevD.99.083008,Nelson_2019,PhysRevD.102.063028}. Further details on the effects on the structure of neutron stars and the electromagnetic observations by this specific DM model can be found in \citep{Miao_2022}. Also, the equations that relate pressure to energy density can be found in section 2 of their paper.

\subsection{Bosonic Dark Matter}\label{sec:bos}
An alternative assumption is that DM is entirely composed of bosonic matter. In this work, the assumption is adapted that DM is described by a massive (mass equals to $m_\chi$) complex scalar field, $\phi$, with self-interaction (with dimensionless coupling constant $\lambda$) \citep{PhysRevLett.57.2485}. The Lagrangian density is then given as
\begin{equation}
    \mathcal{L}=\frac{1}{2}\partial_\mu \phi^*\partial^\mu \phi - \frac{m^{2}_\chi}{2}\phi^* \phi - \frac{\lambda}{4}(\phi^* \phi)^2.
\end{equation}
After applying the mean-field approximation and performing some other computations \citep{PhysRevD.105.023001}, it can be shown that the bosonic DM pressure, P, at a given energy density value, $\epsilon$, is given as
\begin{equation}
    P=\frac{m^4_\chi}{9 \lambda}(\sqrt{1+\frac{3\lambda}{m^4_\chi}\epsilon}-1)^2
\end{equation}

\subsection{Neutron Star's EOS}
In this work, the APR EOS \citep{1998Akmal} is employed. This takes into account 3-nucleon interactions and special relativistic corrections. Other EOS options could also be considered for completeness. However, the focus of this study is on examining how the universal relations are influenced in the presence of DM, and consequently, the choice of the NS's EOS becomes less significant.

\section{Two-fluid DMANS Structure Equations}
\label{sec:2fluid}

\subsection{Two-fluid Tolman-Oppenheimer-Volkoff (TOV) equations}
\label{sec:2TOV}
The simplest model that can be employed to describe two fluids in hydrostatic equilibrium is to neglect interactions among them. This results in the total pressure and energy density of the system being expressed as $P_{tot}(r)=P_D(r)+P_B(r)$ and $\epsilon_{tot}(r)=\epsilon_D(r)+\epsilon_B(r)$. Hereafter, subscripts "D" and "B" will be used to differentiate dark matter from baryonic matter. Consequently, the total energy-momentum tensor can be written as 
\begin{equation}
    T_{tot}^{\mu \nu}=T_D^{\mu \nu}+T_B^{\mu \nu}.
    \label{EN_MOM}
\end{equation}
Since there is no exchange of energy between the two fluids, the continuity equation $\nabla_{\mu}T_{tot}^{\mu \nu}=0$ will be satisfied only if $\nabla_{\mu}T_B^{\mu \nu}=0$ and $\nabla_{\mu}T_D^{\mu \nu}=0$, where $\nabla_{\mu}$ represents the covariant derivative. This leads to two hydrostatic equilibrium conditions, and as a result, the two-fluid Tolman-Oppenheimer-Volkoff (TOV) equations \citep{10.1143/PTP.47.444,SANDIN2009278,CIARCELLUTI201119} are presented as follows

\begin{equation}
\frac{dP_B}{dr} = -\frac{(\epsilon_B(r) +P_B(r))(M_{tot}(r)+4\pi r^{3}P_{tot}(r))}{r[r-2M_{tot}(r)]},
\label{eq:11}
\end{equation}

\begin{equation}
\frac{dP_D}{dr} = -\frac{(\epsilon_D(r) +P_D(r))(M_{tot}(r)+4\pi r^{3}P_{tot}(r))}{r[r-2M_{tot}(r)]},
\label{eq:12}
\end{equation}

\begin{equation}
\frac{dM_B(r)}{dr} = 4\pi r^{2}\epsilon_B(r),
\label{eq:13}
\end{equation}

\begin{equation}
\frac{dM_D(r)}{dr} = 4\pi r^{2}\epsilon_D(r),
\label{eq:14}
\end{equation}
where the total mass is $M_{tot}(r)=M_B(r)+M_D(r)$.

\subsection{Two-fluid Rotating DMANS}
\label{sec:2rns}
In order to model the rotating DMANS a time-independent axisymmetric metric is implemented as follows
\begin{equation}
\begin{aligned}
ds^{2} = -e^{\gamma+\rho}dt^{2} + e^{2\alpha}(dr^{2} + r^{2}d\theta^{2})
\\+ e^{\gamma-\rho}r^{2}\sin^{2}\theta(d\phi-\omega dt)^{2},
\end{aligned}
\end{equation}
where, $\rho$, $\gamma$, $\alpha$, and $\omega$ are referred to as metric potentials, and they depend on r and $\theta$. Note that the coordinate r in this section differs from the one employed in Subsection \ref{sec:2TOV}.

By using the above metric and the two-fluid energy-momentum tensor (equation \ref{EN_MOM}) the hydrostatic equilibrium and the structural equations are derived in the Appendix section \ref{appendix:a}. 

The coordinate radii at the equator of the baryonic and dark matter fluids, $r_{eB}$ and $r_{eD}$, are determined at the points where equations \ref{hydr1} and \ref{eqn:HEDM} are satisfied, respectively.

The circumferential radii at the equator for the two fluids are defined as
\begin{equation}
R_{Be}=r_{Be}e^{(\gamma_{Be}-\rho_{Be})/2}
\end{equation}
for the baryonic fluid, and

\begin{equation}
R_{De}=r_{De}e^{(\gamma_{De}-\rho_{De})/2}
\end{equation}
for the dark matter fluid.

The circumferential polar radius of the baryonic fluid is
\begin{equation}
R_{Bp}=r_{Bp}e^{(\gamma_{Bp}-\rho_{Bp})/2}.
\end{equation}
The subscripts $p$ and $e$ denote the parameter values at the pole and the equator of the fluid, respectively.

The total mass is defined as $M_{tot} \equiv  M_{GD} + M_{GB}$
where $M_{GD}$ and $M_{GB}$ are calculated by equations \ref{eq:Mgb} and \ref{eq:Mgd}, respectively.

Although that it is true that in the non-rotating case, $M_{tot} = M_{GD} + M_{GB} = M_{D} + M_{B}$, where $M_{D}$ and $M_{B}$ are the total masses of the baryonic and DM fluids respectively, computed by using equations \ref{eq:12} and \ref{eq:13}, this is not true for the two fluids separately, as $M_{GB} \neq M_{B}$ and $M_{GD} \neq M_{D}$. For this reason, in this paper the dark matter fraction at the non-rotating limit will be defined as, $f_\chi \equiv M_{D}/(M_{D} + M_{B})$.

\subsubsection{Computational Methods}
To numerically solve the equations, the \texttt{rns} program was modified.
The methods employed for solving the equations remain identical to those mentioned at the beginning of Section \ref{sec:Univ}. The modifications primarily pertain to the metric potential differential equations (\ref{metr1}, \ref{metr2} and \ref{metr3}) and the inclusion of an additional fluid that satisfies the hydrostatic equilibrium equation as indicated above in Eq. (\ref{eqn:HEDM}). The updated value for the equatorial radius of dark matter ($r_{eD}$) during the iterative process is calculated as follows

\begin{equation}
r_{eD} = r'_{eD} \frac{2[h_{Dcenter}( P_{Dcenter})-h_{Dp}]}{[\gamma'_{Dp}+\rho'_{Dp}-\gamma'_{center}-\rho'_{center}]}.
\end{equation}

Here, the symbol "$'$" denotes the previous value of the parameter.

To execute the \texttt{rns} program, it is essential to predefine the ratio of polar to equatorial radii, denoted as $r_{ratio} \equiv r_e/r_p$. Given a specific value of this ratio, and the central density, \texttt{rns} computes the star's spin frequency. In the case of two-fluid stars, two such ratios are defined, $r_{ratioB} \equiv r_{Be}/r_{Bp}$ and $r_{ratioD} \equiv r_{De}/r_{Dp}$, corresponding to the baryonic and dark matter stars that are spinning with different frequencies. Since the baryonic and dark matter are only coupled by gravity, it is expected that the dark matter component have a spin frequency close to zero. For this reason, $r_{ratioD}$ is chosen to be $\approx 1$, so the dark matter fluid will be close to zero spin frequency. This is an approximation, since if there is a non-zero baryonic spin frequency, the spacetime will not be spherically symmetric and therefore $r_{ratioD} \neq 1$.

\subsubsection{Accuracy Tests in the Non-rotating Limit}

The data describing the star's properties in the non-rotating case are compared to the two-fluid TOV equation solutions from Subsection \ref{sec:2TOV}, in order to test the accuracy of the code.

The fractional difference in the values of the total mass and outer radius of the DANS computed using the two methods is typically less than 0.25\% and 0.48\% respectively. These differences mainly come about from the different methods used to determine the location of the baryonic surface of the DANS. The magnitude of these errors are similar to those encountered when using rns for pure neutron stars.

\section{Shape-Related Universal Relations}
\label{sec:Univ}

Several universal relations have emerged in various studies of rotating neutron stars, showing relative independence of many bulk properties from the neutron star's EOS. A few examples are the dimensionless moment of inertia - compactness relation \citep{1994ApJ...424..846Ravenhall}, the dimensionless gravitational binding energy - compactness relation \citep{2001ApJ...550..426Lattimer}, the relation among the maximum mass of the non-rotating and rotating neutron stars \citep{1996ApJ...456..300Lasota,2016MNRAS.459..646Breu}, and the relation among the dimensionless moment of inertia, the quadrupole moment and the Love number, also known as the ``I-Love-Q" relations \citep{2013Yagi,2017PhR...681....1Yagi}. 
The effect of dark matter on the I-Love-Q relations has recently been investigated (\citep{2023arXiv230907971W}, \citep{2023PhRvD.108j3016C}).

In this paper, the focus is on changes caused by dark matter to the universal relations of the oblate shape of neutron stars due to rotation \citep{2007Morsink,2021PhRvD.103f3038Silva}, and the changes in mass and radius due to rotation in sequences of neutron stars with constant central energy density \citep{Konstantinou_2022}.

\subsection{Universal Relations for the Increase in the Mass and Radius of a Rotating Neutron Star}

In previous work \citep{Konstantinou_2022}, it was demonstrated that the change in mass and equatorial radius of rotating NSs is quasi-independent of the choice of the  EOS, when comparing between rotating and non-rotating stars with the same central energy density.

\citet{Konstantinou_2022} introduced two sets of empirical equations to describe this universality. The first set can be used to construct the mass-radius curve for rotating stars, starting from the non-rotating TOV solution. The second set, utilized in this paper, can be employed to reconstruct the mass-radius curve for non-rotating stars, given measurements of mass, radius, and spin for rotating neutron stars. The latter set is as follows.
Given a rotating star's mass, $M$, equatorial radius, $R_e$, and spin frequency $\nu$, the mass and radius $M_*$ and $R_*$ of a non-rotating star with the same central density as the rotating star are given by the formulae
\begin{equation}
\begin{aligned}
\frac{R_e}{R_*} =  1 +  ( e^{A_{2} \Omega_{n2}^2} - 1 + B_{2} \left[ \ln( 1 - (\frac{\Omega_{n2}}{1.1})^4)\right]^2)
\\ \times \left( 1 + \sum_{i=1}^{5} b_{r,i} C_e^i \right),
\end{aligned}
\label{eq:Rinvfit}
\end{equation}
and 
\begin{equation}
\frac{M}{M_*} = 1 + \left( \sum_{i=1}^4 d_i \Omega_{n2}^i\right)  \times \left( \sum_{i=1}^{4} b_{m,i} C_e^i \right),
\label{eq:Minvfit}
\end{equation}
where $C_e$ is the equatorial compactness $\frac{M}{R_e}$, and $\Omega_{n_2}$ denotes the normalized spin frequency of the star (that depends only on $\nu$, $M$, and $R_e$) and the values of coefficients $A_2$, $B_2$, $b_{r,2}$, and $b_{m,i}$ are provided in Table~1 of \citep{Konstantinou_2022}.
These formulae are universal in the sense that given an EOS, all rotating neutron stars constructed with that EOS are mapped to the same zero-spin mass-radius curve, allowing for comparisons of mass and radius measurements for neutron stars with different spin frequencies.

\subsection{Oblate Shape Universality}

\citet{2007Morsink} have highlighted that the change in the shape of neutron stars due to rotation is universal. They presented an empirical equation (see their Equation 8) that relates the radius at an angle $\theta$ from the rotational axis, $R(\theta)$, to the equatorial radius, $R_e$. This equation depends on several coefficients  as well as the dimensionless compactness $\kappa$ and the dimensionless squared frequency $\sigma$, defined as
\begin{equation}
    \kappa \equiv \frac{GM}{R_{e}c^2}, \quad
    \sigma \equiv \frac{\Omega^2 R^3_{e}}{GM}.
\end{equation}
The universal oblate shape function is used extensively to infer the masses and radii of rotating pulsars through pulse-profile modelling \citep{2019bBogdanov}, so it is important to understand how dark matter affects the oblate shape. 

\cite{2021PhRvD.103f3038Silva} introduced an improved version of the empirical shape, the ellipsoidal model, which provides better predictions when the star rotates rapidly ($\sigma > 0.2$). When $\sigma = 0$, they enforce $R(\theta)$ to be equal to $R_e$ (the \citet{2007Morsink} equation is characterized by a mismatch of 1\%). 

In this paper, the ellipsoidal formula is utilized and is defined as follows \citep{2021PhRvD.103f3038Silva}
\begin{equation}
    R(\theta)=R_{e}\sqrt{\frac{1-e^2}{1-e^2 g(cos(\theta))}},
    \label{eq:shape}
\end{equation}
where e and $g(cos(\theta))$ are functions dependent on $\kappa$ and $\sigma$, along with a set of numerical coefficients (see their Equations (17), (19) and Table 3). Specifically, there are two sets of these coefficients: one for the slow rotation limit and one for the fast rotation limit. In this paper, the former limit is employed when $\sigma \leq 0.2$, and the latter is used when $\sigma>0.2$

\section{Results}\label{sec:non-universal}
\subsection{Universal Relations for a Pure Neutron Star}

\begin{figure}
    \centering
    \includegraphics[width=90mm]{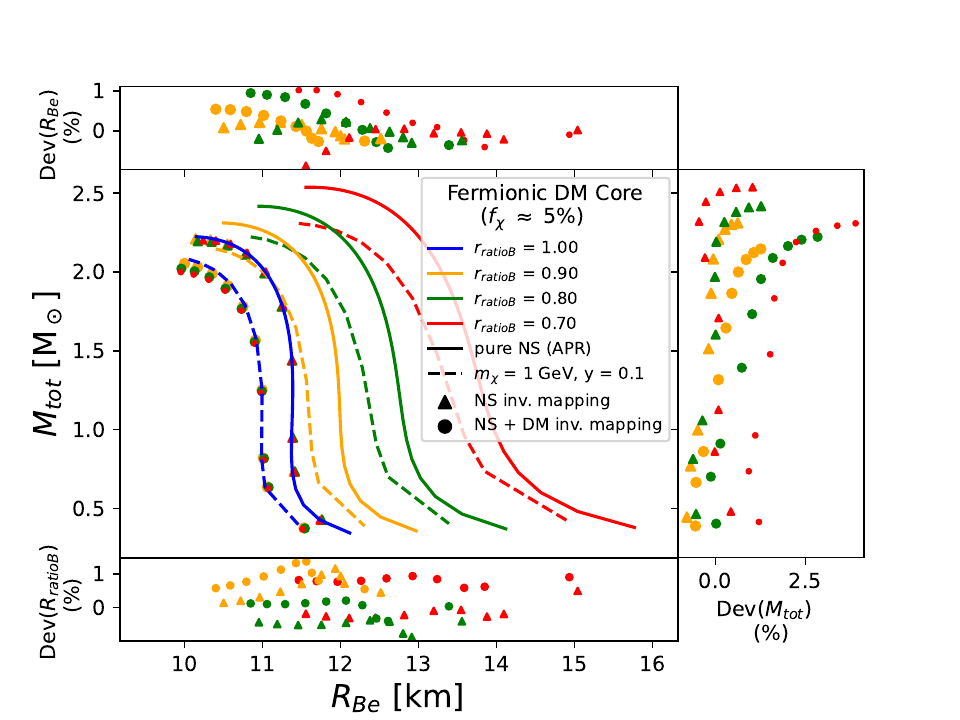}
    \includegraphics[width=90mm]{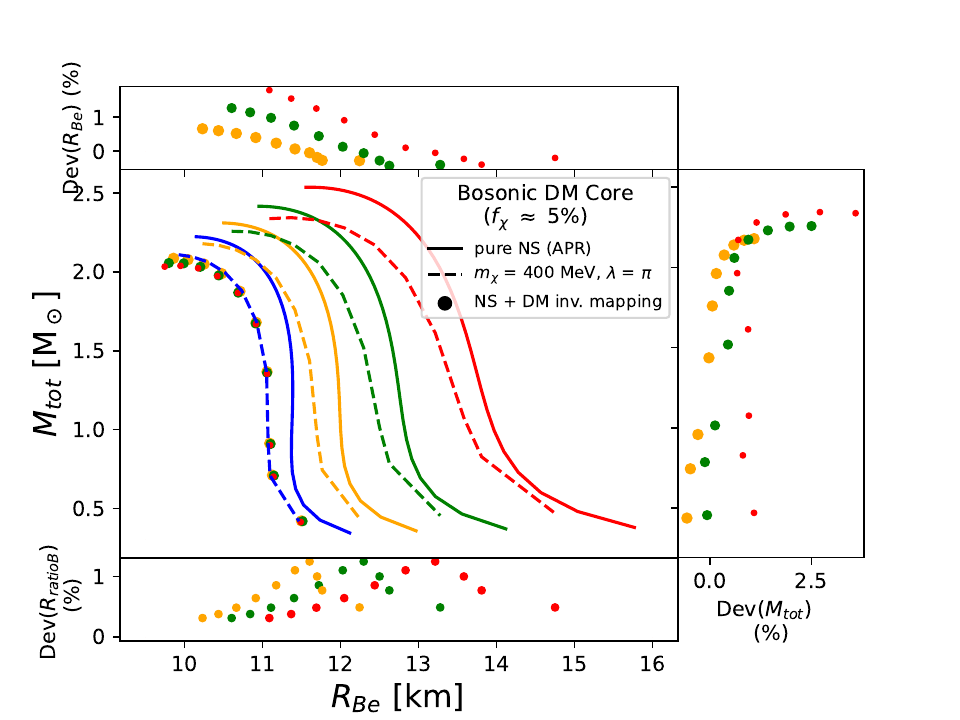}
    \caption{This figure illustrates the relationship between the total mass, $M_{tot}$, and the baryonic equatorial radius of rotating neutron stars under two cases: those composed of pure nuclear matter (indicated by solid curves, using the APR EOS), and those comprising rotating DMANS with approximately a 5\% DM mass fraction (indicated by dashed curves, featuring a fermionic (upper) or a bosonic (lower) DM core). Distinct colors correspond to various $r_{ratioB}$ values: blue represents 1.0, orange for 0.9, green for 0.8, and red for 0.7. The figure also includes circular and triangular points that signify the inverse mapping conducted from the rotating stars to the non-rotating ones. Triangular points denote scenarios with pure nuclear matter, while circular points signify cases with the presence of DM. The plot right and above the $M_{tot}$ vs $R_{Be}$ plot, show the percent deviation of the reproduced data compared to the predictions derived from equations \ref{eq:Rinvfit} and \ref{eq:Minvfit}, represented as $Dev(R_{Be})$ and $Dev(M_{tot})$, respectively. At the bottom, $|Dev(R_{ratioB})|$ is illustrated.}
    \label{fig:un}
\end{figure}

\begin{figure}[ht]
    \centering
     \includegraphics[width=80mm]{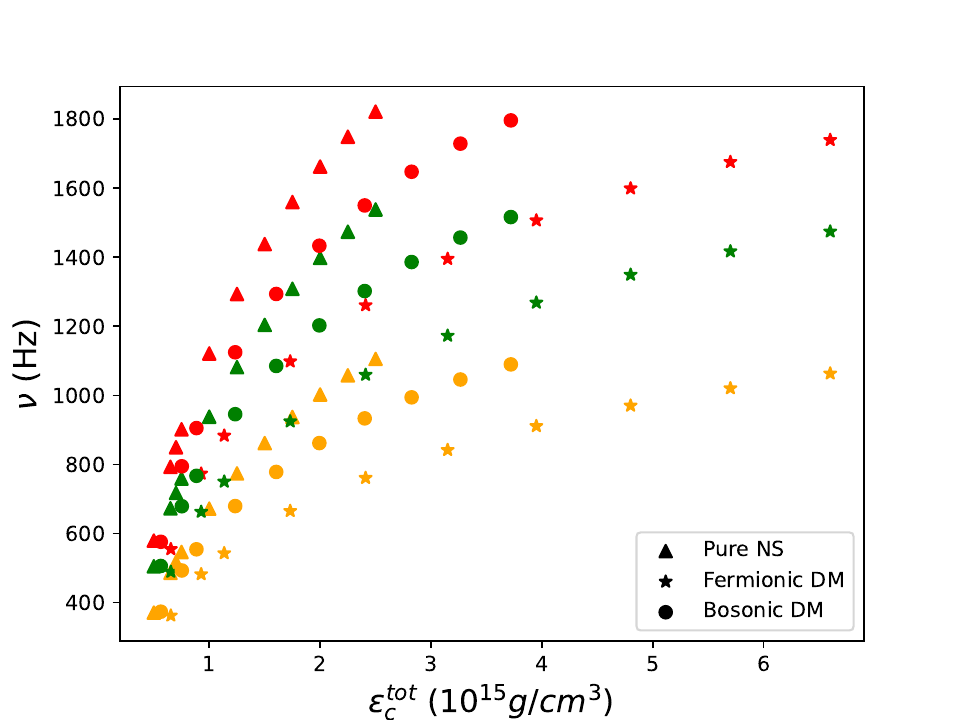}
    \caption{ Spin frequency vs central total energy density of the stars  presented in Figure \ref{fig:un}. The triangle, star and circular shaped points, represent the pure, Fermionic DM admixed and Bosonic DM admixed  NSs, respectively. The stars with the largest central total energy density values represent the stars with the largest mass values, and vice versa.
 }
    \label{fig:freq}
\end{figure}

The application of the universal relations from Section \ref{sec:Univ} to a pure neutron star (NS) is demonstrated in Figure \ref{fig:un}. Figure \ref{fig:un} presents four mass-radius curves (solid lines) for EOS APR generated using the \texttt{rns} program, for four different values of $r_{ratio}$. The $r_{ratio}$ values used are 1.0 (blue), 0.9 (orange), 0.8 (green), and 0.7 (red). The neutron stars on the blue curve are not spinning, while the spin frequencies the stars on the orange, green, and red curves span a wide range of frequencies from 400 - 1800 Hz, as shown by the triangle symbols in Figure~\ref{fig:freq}.

The universality of the changes in mass and radius of a neutron star caused by spin can be tested by taking any point on one of the orange, green, or red solid curves, and using the mass, radius, and spin frequency in equations (\ref{eq:Minvfit}) and (\ref{eq:Rinvfit}) to predict values $M_*$ and $R_*$ for the non-rotating star with the same EOS and central density. These values are plotted in Figure 1 (upper panel) as triangles. The non-rotating mass-radius curve is accurately reconstructed by using points from the Mass-Radius curves with $r_p/r_e$ equal to 0.9, 0.8, and 0.7. The triangles of different colours are mapped to locations on the blue (non-rotating) curve, independent of the rotation rate of the rotating star, which is the expression of universality. 

More quantitatively, the percent deviation of a quantity $Q$ from its corresponding best-fit predicted value $Q_{fit}$ is defined by
\begin{equation}
 Dev(Q) = 100 \times \frac{Q-Q_{fit}}{Q}.
\end{equation}
The percent deviations in the radius arising from Equation (\ref{eq:Rinvfit}) are shown in the panel above the mass-radius curves in Figure \ref{fig:un} with triangles. The percent deviations in the radius are less than 1\% for the pure neutron stars. Similarly, the percent deviations in the mass coming from the universal formula (\ref{eq:Minvfit}), shown with triangles to the right of the mass-radius curves in Figure \ref{fig:un} are also mostly less than 1\%, except for the most massive and rapidly rotating neutron stars. 

The universality of the ellipsoidal shape function (\ref{eq:shape}) is tested through the percent deviation of the change in $R_{ratioB} \equiv R_Bp/R_Be$ of a pure neutron star, Dev($R_{ratioB}$). The deviations in the polar-to-equatorial ratio resulting from the ellipsoidal equation (evaluated at $\theta = 0$) are shown below the mass-radius curves with triangles and are $\lessapprox 1.25 \%$. The values of predicted and calculated radius at other values of angle were also computed and corresponded to similar or smaller errors. 

The small values of the percent deviations in the mass, radius and shape from the empirical formulae means that these formulae are universal for pure neutron stars.

\subsection{Rotating Dark Matter Admixed Neutron Stars}

The investigation now proceeds to explore the impact of DM core on the universal relations. The total masses, equatorial and polar radii of DMANS for two different DM equations of state (one fermionic and the other bosonic) and the baryonic EOS APR are calculated using the assumptions and the two-fluid rns code detailed in Sections \ref{sec:EOS} and \ref{sec:2rns}.  In each subcase, the free parameter (central dark matter density) is chosen so that a DM core is formed. 

Mass-radius curves for non-rotating DMANS with a dark matter mass fraction $f_\chi=0.05$ are shown in Figure \ref{fig:un} with blue dashed curves for a fermionic DM EOS (upper panel) and a bosonic DM EOS (lower panel). Since these are DMANS models with DM cores, the mass in this case is the total mass contained within the DMANS, and the radius is the outer radius of the baryonic matter. 
As is typical for DM cores \citep{Miao_2022}, the addition of DM softens the overall EOS, resulting in DMANS with smaller masses and radii than pure neutron stars with the same baryonic EOS, as can be seen by comparing the blue solid and dashed curves in either the upper or lower panels. The radii of the DM core range from about 30 - 50\% (50 - 90\%) of the fermionic (bosonic) radius. 

Mass-radius curves for rotating DMANS with fixed values of $r_{ratio}$ are shown in Figure \ref{fig:un} with dashed lines with the same colour choices as for the pure neutron stars. The central energy densities for each rotating DMANS were chosen to match the values of a non-rotating DMANS. This procedure does not preserve the value of $f_\chi$.
\begin{figure}
    \centering    \includegraphics[width=80mm]{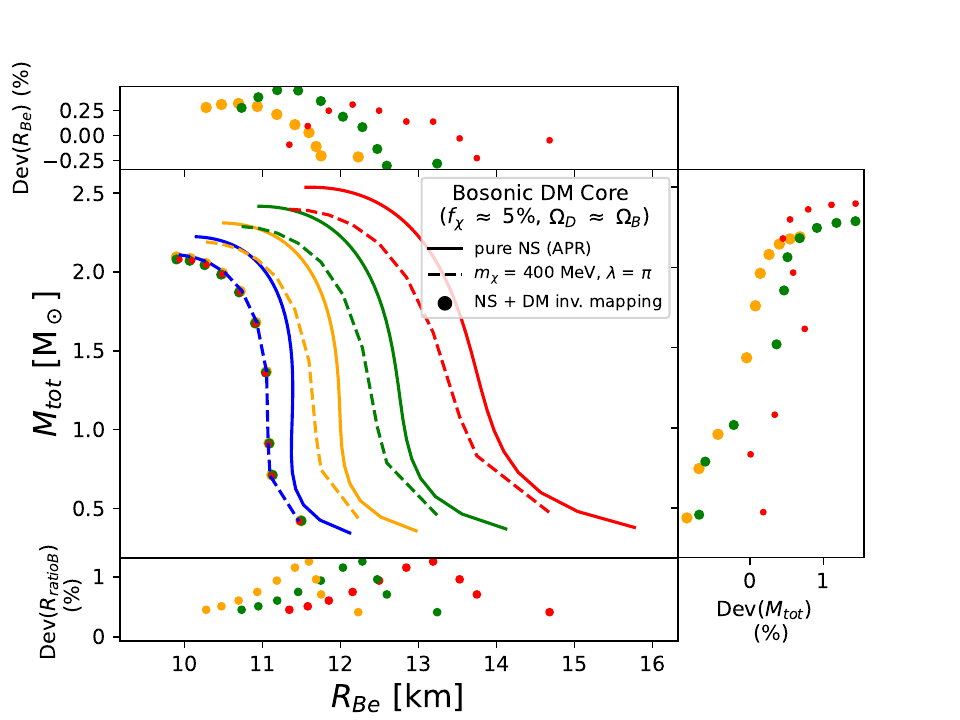}
    \caption{Same as the bosonic case in Figure \ref{fig:un}, but now the DM and the baryonic fluids are rotating with the same angular velocity. }
    \label{fig:1freq}
\end{figure}

The inverse mappings between rotating and non-rotating masses and radii are calculated using universal equations \ref{eq:Rinvfit} and \ref{eq:Minvfit} for the rotating DMANS and shown in Figure \ref{fig:un} as colored circular points. These points come close to overlapping the dashed blue curves representing the zero spin DMANS, illustrating that the universal equations do not appear to be strongly affected by the presence of a dark matter core. There appear to be slightly larger errors, near the maximum mass stars, when dark matter is present.
 However, the deviations are small, even in the presence of a DM core ($|Dev(R_e)| \lessapprox 1.8 \%$ and $|Dev(M_{tot})| \lessapprox 3.9 \%$), which is illustrated in the panels above and the right of the mass-radius curves shown in Figure \ref{fig:un} for both fermionic and bosonic dark matter models. 
 
 At the bottom of Figure \ref{fig:un} the percent deviation of $R_{ratioB}$ from the empirical universal equation \ref{eq:shape} is depicted. Although $Dev(R_{ratioB})$ becomes slightly larger with a DM core, this change remains within $|Dev(R_{ratioB})| \lessapprox 1.4\%$, indicating that universality is maintained. Similar to the pure neutron star case, the shape universality represented by equation \ref{eq:shape} holds for all the possible values of $\theta$ (not only for $\theta$ = 0).

\begin{figure}[H]
    \centering
    \includegraphics[width=80mm]{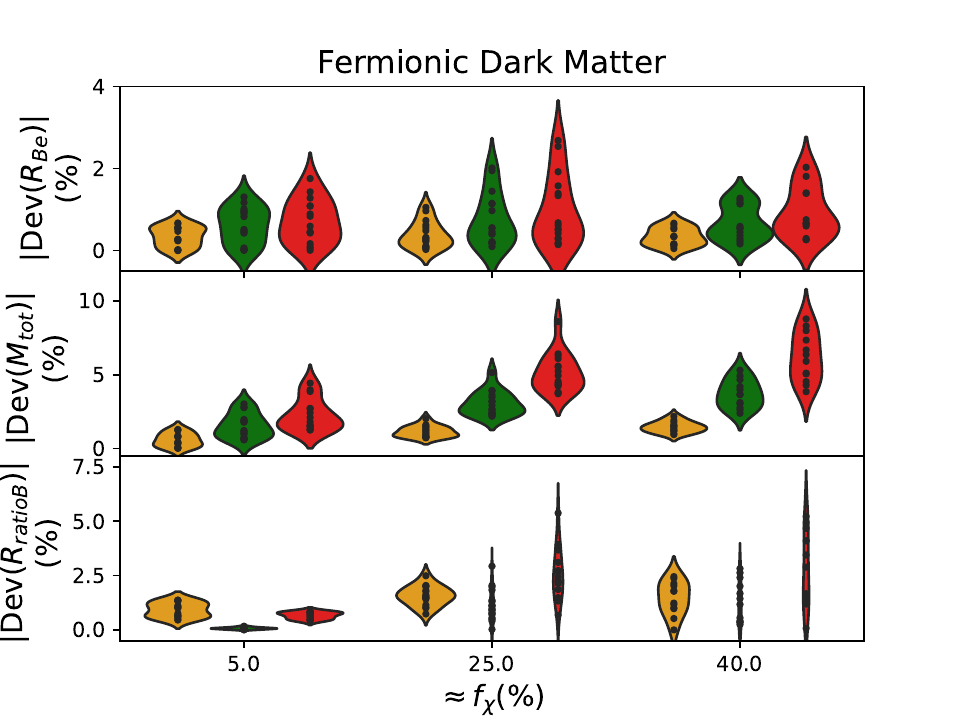}
    \includegraphics[width=80mm]{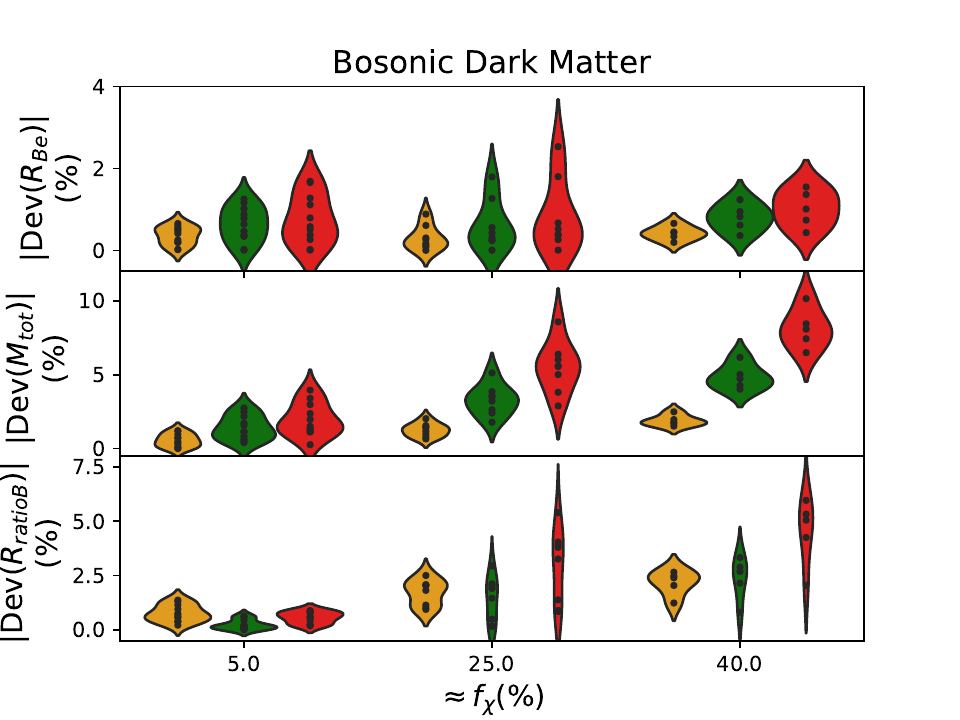}
    \caption{A violin plot is displayed, showing the percent deviation of the reproduced data with different $f_\chi$ and free parameter values, compared to the predictions derived from the universal equations for changes in $M_{tot}$, $R_{Be}$, and $R_{ratioB}$, represented as $Dev(R_{Be})$, $Dev(M_{tot})$, and $|Dev(R_{ratioB})|$, respectively. 
The black dots within the violin plot represent individual data points used in the analysis. The curved shape of the violin plot is generated from a kernel density estimation, which provides a smooth representation of the underlying data distribution.  }
    \label{fig:Parameters}
\end{figure}
\subsection{Dependence of Universal Relations on Free Parameters}

The difference between the deviations between the fermionic core case and the bosonic core case is insignificant. This suggests that the choice of DM EOS is unimportant, as long as a DM core forms (as opposed to a halo). The choice of $f_\chi = 0.05$ is a fairly large dark matter fraction. If smaller (and probably more likely) dark matter fractions were used, the deviations introduced by the empirical formulae would be smaller. 

Furthermore, the examination of the dependence of the universal relations on the choice of the values of the free parameters has been conducted. The values of the parameters that generate a DM core with $f_\chi$ = 5\% are given in Figure 3 (Figure 10) of \citet{Miao_2022} \citep{PhysRevD.109.043029} for fermionic (bosonic) DM. Three combinations of parameters have been chosen for each case to cover different regions of the parameter space. Figure \ref{fig:Parameters} illustrates the deviation of the produced data from the three universal relations introduced in this paper. The left-hand side shows the case where fermionic DM is used with [$m_\chi$ = 12 GeV, y = 1000], [$m_\chi$ = 0.6 GeV, y = 0.01], [$m_\chi$ = 1 GeV, y = 1], and the right-hand side shows the case where bosonic DM is used with [$m_\chi$ = 300 MeV, $\lambda$ = 0.1 $\pi$], [$m_\chi$ = 250 MeV, $\lambda$ = 5 $\pi$], [$m_\chi$ = 150 MeV, $\lambda$ = 0.2 $\pi$]. In addition, the fermionic DM plot includes a mirror DM DMANS, where the DM has the same EOS as baryonic matter, but doesn't interact with ordinary matter  \citep{SANDIN2009278, CIARCELLUTI201119,Goldman_2011,GOLDMAN2013200, 2021PhRvD.103d3009K}. Four different baryonic central energy density values have been chosen ($1.0 \times 10^{15}$ g $cm^{-3}$, $1.5 \times 10^{15}$ g $cm^{-3}$, $2.0 \times 10^{15}$ g $cm^{-3}$, $2.5 \times 10^{15}$ g $cm^{-3}$), while the DM central energy density was varied accordingly to generate DMANS with $f_\chi \approx$ 5\%. 
Similarly, DMANS with $f_\chi \approx$ 25\% and $f_\chi \approx$ 40\% were also computed as extreme (if physically unlikely) cases in order to test if the universality could be broken. 
The same color-scheme as Figure \ref{fig:un} has been utilized to separate the three different $r_{ratio}$ choices (orange for 0.9, green for 0.8, and red for 0.7).

The similarity of the left and right panels of Figure \ref{fig:Parameters} illustrates that the universal relations are affected in the same manner by fermionic and bosonic DM EOS. Furthermore, the DMANS with $f_\chi \approx$ 5\% are characterized by a similar deviation as the cases illustrated in Figure \ref{fig:un}. Therefore, it can be concluded that the universal relations are not affected by the choice of the free parameters. The largest increases in the deviations can be achieved at $r_{ratio} = 0.7$ and by adding more dark matter to the system. 
However, most pulsars observed by NICER are described by models with $r_{ratio} \le 0.9$ and it is unlikely that DMANS with dark matter fractions of 25\% or higher exist.


An extreme (if unphysical) case was constructed in order to further test the universal relations by computing DMANS where the two fluids are spinning with the same frequency. This case is shown in Figure \ref{fig:1freq}, with the corresponding percent deviations to be $|Dev(R_e)| \lessapprox 0.31 \%$, $|Dev(M_{tot})| \lessapprox 1.44 \%$ and $|Dev(R_{ratioB})| \lessapprox 1.26 \%$. It can be seen that the maximum values of $Dev(R_e)$, $Dev(M_{tot})$ and $Dev(R_{ratioB})$ decrease.

\section{Discussion}
\label{sec:discussion}
Neutron stars can provide insights into the properties of supranuclear density matter and have the potential to reveal characteristics of dark matter if they accrete a sufficient amount of it. This paper has focused on an examination of how universal relations pertaining to the shape of rapidly rotating neutron stars are altered in the presence of dark matter.

To investigate this, the \texttt{rns} program was modified under the assumption that the dark matter and baryonic matter can be approximated as two perfect fluids, non-interacting, cold, and rigidly rotating with aligned axes. The dark matter fluid spin frequency is chosen to be zero, since it is hypothesized that this would represent a most realistic case.

For the modeling of dark matter, two pre-existing phenomenological dark matter EOSs (one fermionic and one bosonic), were utilized. Each of these models is characterized by two free parameters governing particle mass and interaction strength. Calculations were carried out for scenarios in which the dark matter formed a dark matter core contained inside of the baryonic surface of the star.

The findings indicate that the universal relations concerning changes in mass and equatorial radius, as well as shape universality, remain unaltered by the presence of a dark matter core. 
\section{Acknowledgements}
 I'd like to express my gratitude to Sharon Morsink and Shafayat Shawqi for valuable comments, suggestions, and discussions. Additionally, I extend my thanks to Anastasios (Tasos) Irakleous, Ilya Mandel and Paul Lasky for our insightful conversations. I also acknowledge the use of GPT-3.5, the AI language model developed by OpenAI, for its assistance in identifying and rectifying any grammatical or syntactical errors.
  \newpage
\appendix
\section{Appendix: Detailed Two-fluid Rotating DMANS model}
\label{appendix:a}
This section delves into the subject of two-fluid stationary, axisymmetric, rotating compact stars. Therefore, the two fluids are assumed to rotate along the same axis. 

The time-independent axisymmetric metric is given as follows
\begin{equation}
ds^{2} = -e^{\gamma+\rho}dt^{2} + e^{2\alpha}(dr^{2} + r^{2}d\theta^{2}) + e^{\gamma-\rho}r^{2}\sin^{2}\theta(d\phi-\omega dt)^{2},
\end{equation}
where, $\rho$, $\gamma$, $\alpha$, and $\omega$ are referred to as metric potentials, and they depend on r and $\theta$. Note that the coordinate r in this section differs from the one employed in Section \ref{sec:2fluid}.

The two fluids are assumed to be ideal and non-interacting. Hence, the total energy-momentum tensor is defined as 
\begin{equation}
T_{tot}^{\mu \nu}=T_{D}^{\mu \nu}+T_{B}^{\mu \nu},
\end{equation}
where $T_{B}^{\mu \nu}$ is the baryonic matter energy-momentum tensor
\begin{equation}
T_{B}^{\mu \nu} = (\epsilon_B + P_B)u_B^{\mu}u_B^{\nu}+( P_B)g^{\mu \nu},
\end{equation}
with the baryonic four-velocity, $u_B^{\nu}$ to be equal to
\begin{equation}
u_B^\alpha = \frac{e^{-(\gamma+\rho)/2}}{(1-\upsilon_B^2)^{1/2}}[1,0,0,\Omega_B],
\end{equation}
where
\begin{equation}
\upsilon_B =(\Omega_B-\omega)r\sin \theta e^{-\rho}.
\end{equation}
$\Omega_B$ is the angular velocity of the baryonic fluid. 

Similarly, for the DM fluid we have that
\begin{equation}
T_{D}^{\mu \nu} = (\epsilon_D+ P_D)u_D^{\mu}u_D^{\nu}+( P_D)g^{\mu \nu},
\end{equation}
with
\begin{equation}
u_D^\alpha = \frac{e^{-(\gamma+\rho)/2}}{(1-\upsilon_D^2)^{1/2}}[1,0,0,\Omega_D],
\end{equation}
where
\begin{equation}
\upsilon_D =(\Omega_D-\omega)r\sin \theta e^{-\rho}.
\end{equation}
Note that if the dark matter component has zero angular velocity, $\upsilon_D$ will be non-zero due to the frame-dragging potential $\omega$ caused by the rotation of the baryonic component.

The Einstein's field equations are expressed as
\begin{equation}
G^{\mu \nu} \equiv R^{\mu \nu}-\frac{1}{2}g^{\mu \nu} R=8 \pi T_{tot}^{\mu \nu},
\end{equation}

where $G^{\mu \nu}$,  $R^{\mu \nu}$ and $R$ represent the Einstein tensor, the Ricci tensor and the Ricci scalar, respectively. This equation is analogous to the one-fluid case \citep{1971ApJ...167..359B}, with the sole distinction that $T^{\mu \nu}$ is replaced with $T_{B}^{\mu \nu}+T_{D}^{\mu \nu}$. 

One can show that
\begin{equation}
G^{1 1} = r^2 G^{2 2} = 8 \pi e^{-2\alpha} (P_D + P_B),
\end{equation}

\begin{equation}
G^{1 2} = G^{2 1} = 0,
\end{equation}

\begin{equation}
 G^{0 3}-\omega G^{0 0}=  8 \pi e^{-\gamma-\rho} \Bigr[(\epsilon_B+P_B)\frac{\Omega_B-\omega}{1-\upsilon_B^2} + (\epsilon_D
+P_D)\frac{\Omega_D-\omega}{1-\upsilon_D^2} \Bigr],
\end{equation}

\begin{equation}
F \equiv (\omega^2 G^{0 0}-2\omega G^{0 3}+G^{3 3})e^{-2\rho}sin^2(\theta) r^2 + G^{0 0}=  8 \pi e^{-\gamma-\rho} \Bigr[(\epsilon_B+P_B)\frac{1+\upsilon_B^2}{1-\upsilon_B^2} + (\epsilon_D
+P_D)\frac{1+\upsilon_D^2}{1-\upsilon_D^2} \Bigr],
\end{equation}
\begin{equation}
F - G^{1 1} e^{2\alpha} =  8 \pi e^{-\gamma-\rho} \Bigr[\frac{\epsilon_B(1+\upsilon_B^2)+2P_B\upsilon_B^2}{1-\upsilon_B^2} + \frac{\epsilon_D(1+\upsilon_D^2)+2P_D\upsilon_D^2}{1-\upsilon_D^2} \Bigr],
\end{equation}

Consequently, the structure of the metric potential differential equations (equations 10, 11, and 12 in \citep{1989Komatsu}) remain the same, with the only difference that now
\begin{equation}
\begin{aligned}
S_\rho(r,\mu) = e^{\gamma/2}\Biggr[8 \pi e^{2\alpha}(\epsilon_B+P_B)\frac{1+\upsilon_B^2}{1-\upsilon_B^2} + 8 \pi e^{2\alpha}(\epsilon_D
+P_D)\frac{1+\upsilon_D^2}{1-\upsilon_D^2}  + r^2(1-\mu^2)e^{-2\rho} \Bigr[ \omega_{,r}^2 + \frac{1}{r^2}(1-\mu^2)\omega_{,\mu}^2 \Bigr] 
\\ + \gamma_{,r}/r - \frac{1}{r^2} \mu \gamma_{,\mu} + \frac{\rho}{2} \Biggl\{ 16 \pi e^{2\alpha} (P_D + P_B) - \gamma_{,r} (\frac{1}{2} \gamma_{,r} + \frac{1}{r}) 
- \gamma_{,\mu}/r^2 \Bigl[\frac{1}{2} \gamma_{,\mu} (1-\mu^2) - \mu \Bigr]  \Biggr\} \Biggr]
\label{metr1}
\end{aligned}
\end{equation}
\begin{equation}
\begin{aligned}
S_\gamma(r,\mu) = e^{\gamma/2} \Biggr[ 16 \pi e^{2\alpha} (P_D + P_B)+ \frac{\gamma}{2} \Biggr\{ 16 \pi e^{2\alpha} (P_D 
+ P_B) - 1\frac{1}{2}\gamma_{,r}^2 - 1\frac{1}{2r^2} (1-\mu^2)\gamma_{,\mu}^2)\Biggr\} \Biggr] 
\label{metr2}
\end{aligned}
\end{equation}

\begin{equation}
\begin{aligned}
S_\omega(r,\mu) = e^{(\gamma-2\rho)/2}\Biggr[-16 \pi e^{2\alpha}\frac{(\Omega_B-\omega)(\epsilon_B+P_B)}{1-\upsilon_B^2} -16 \pi e^{2\alpha}\frac{(\Omega_D-\omega)(\epsilon_D+P_D)}{1-\upsilon_D^2} + \omega \Biggl\{ -8 \pi e^{2\alpha}\frac{[(1+\upsilon_B^2)\epsilon_B+2\upsilon_B^2 P_B]}{1-\upsilon_B^2} - 
\\8 \pi e^{2\alpha}\frac{[(1+\upsilon_D^2)\epsilon_D+2\upsilon_D^2 P_D]}{1-\upsilon_D^2} - \frac{1}{r}(2\rho_{,r}+\frac{1}{2}\gamma_{,r}) + \frac{1}{r^2}\mu(2\rho_{,\mu}
+\gamma_{,\mu}/2) + \frac{1}{4}(4\rho_{,r}^2-\gamma_{,r}^2)  \Biggr\} + \frac{1}{4r^2}(1-\mu^2)(4\rho_{,\mu}^2-\gamma_{,\mu}^2) 
\\- r^2(1-\mu^2)e^{-2\rho} \Bigr[ \omega_{,r}^2+\frac{1}{r^2}(1-\mu^2)\omega_{,\mu}^2 \Bigr]  \Biggr\} \Biggr]
\label{metr3}
\end{aligned}
\end{equation}
where $\mu = cos(\theta)$, and $_{,\mu}$ and $_{,r}$ are the partial derivatives with respect to $\mu$ and r, respectively.

Next, by assuming rigid rotation for both fluids and applying similar logic as in the continuity equation in section \ref{sec:2TOV}, where $\nabla_{\mu} T_B^{\mu \nu}=0$ and $\nabla_{\mu}T_D^{\mu \nu}=0$, two hydrostatic equilibrium equations can be derived.

One for the baryonic fluid
\begin{equation}
\begin{aligned}
h_B( P_B)-h_{Bp} \equiv \int_{P_{Bp}}^{P_B} \frac{dP_B}{\epsilon_B+P_B}= lnu_B^t-lnu^t_{Bp}
= \biggr[\gamma_{Bp}+\rho_{Bp}-\gamma-\rho-ln(1-\upsilon_B^2)\biggr]/2,
\end{aligned}
\end{equation}
and one for the dark matter fluid
\begin{equation}
\begin{aligned}
h_D(P_D)-h_{Dp} \equiv \int_{P_{Dp}}^{P_D} \frac{dP_D}{\epsilon_D+P_D} = lnu_D^t-lnu^t_{Dp}=
\biggl[\gamma_{Dp}+\rho_{Dp}-\gamma-\rho-ln(1-\upsilon_D^2)\biggr]/2,
\end{aligned}
\end{equation}

where the subscripts $Bp$ and $Dp$ denote the parameter values at the pole of the baryonic and dark matter fluid, respectively.

The coordinate radii at the equator of the baryonic and dark matter fluids, $r_{eB}$ and $r_{eD}$, are determined at the points where the following two relations are satisfied, respectively.
For the baryonic fluid
\begin{equation}
h_{Bcenter}( P_{Bcenter})-h_{Bp} =\frac{1}{2}[\gamma_{Bp}+\rho_{Bp}-\gamma_{center}-
\rho_{center}],
\label{hydr1}
\end{equation}

and for the dark matter fluid
\begin{equation}
h_{Dcenter}( P_{Dcenter})-h_{Dp} =\frac{1}{2}[\gamma_{Dp}+\rho_{Dp}-\gamma_{center}-\rho_{center}].
\label{eqn:HEDM}
\end{equation}

The circumferential radii at the equator for the two fluids are defined as
\begin{equation}
R_{Be}=r_{Be}e^{(\gamma_{Be}-\rho_{Be})/2}
\end{equation}
for the baryonic fluid, and

\begin{equation}
R_{De}=r_{De}e^{(\gamma_{De}-\rho_{De})/2}
\end{equation}
for the dark matter fluid.

The circumferential polar radius of the baryonic fluid is
\begin{equation}
R_{Bp}=r_{Bp}e^{(\gamma_{Bp}-\rho_{Bp})/2}.
\end{equation}
This is the polar radius that has been used to test the surface universality.

Then the total mass of the system is given by
\begin{equation}
 M_{tot} =\int  (-2T^0_0+T^\mu_\mu) \sqrt{-g} dr\: d\theta\: d\phi,
\end{equation}

where $g$ is the determinant of the metric $g_{\mu \nu}$. As a result $M_{tot} =  M_{GD} + M_{GB}$
where 
\begin{equation}
\begin{aligned}
 M_{GB} = \int e^{2\alpha+\gamma}  \{\frac{\epsilon_B+P_B}{1-\upsilon_B^2}[1+\upsilon_B^2+2\omega\: r\:sin\theta\:e^{-\rho}\upsilon_B]
+2P_B\} r^2\: sin\theta\: dr\: d\theta \: d\phi,
\label{eq:Mgb}
 \end{aligned}
\end{equation}
and
\begin{equation}
\begin{aligned}
M_{GD} = \int e^{2\alpha+\gamma}  \{\frac{\epsilon_D+P_D}{1-\upsilon_D^2}[1+\upsilon_D^2+2\omega\: r\:sin\theta\:e^{-\rho}\upsilon_D]
+2P_D\} r^2\: sin\theta\: dr\: d\theta \: d\phi.
\label{eq:Mgd}
\end{aligned}
\end{equation}

\bibliography{biblio}{}
\bibliographystyle{aasjournal}

\listofchanges

\end{document}